\documentclass[letterpaper,10pt]{article}

\usepackage{epsf}
\usepackage{caption}
\usepackage[justification=centering]{subfig}
\usepackage{epsfig}
\usepackage{color}
\usepackage{rotating}
\usepackage{relsize}
\usepackage{array}
\usepackage{multirow}
\usepackage{amsmath}
\usepackage{amssymb}
\usepackage{amsfonts}
\usepackage{mathtools}
\usepackage{times}
\usepackage{datetime}
\usepackage{graphicx}
\usepackage{enumerate}
\usepackage[mathscr]{euscript}
\usepackage{algorithm}
\usepackage{booktabs}
\usepackage{colortbl}
\usepackage{accents}

\usepackage[margin=1.0in]{geometry}
\newcommand{\comments}[1]{}
\newcommand{\R}{\mathbb{R}}

\renewcommand{\(}{\left(}
\renewcommand{\)}{\right)}

\renewcommand{\emptyset}{\varnothing}

\renewcommand{\phi}{\varphi}
\renewcommand{\tt}[1]{\texttt{#1}}

\newcolumntype{L}[1]{>{\raggedright\let\newline\\\arraybackslash\hspace{0pt}}m{#1}}
\newcolumntype{C}[1]{>{\centering\let\newline\\\arraybackslash\hspace{0pt}}m{#1}}
\newcolumntype{R}[1]{>{\raggedleft\let\newline\\\arraybackslash\hspace{0pt}}m{#1}}

\newenvironment{itemize*}
	{\vspace{-6pt} \begin{itemize} \setlength{\itemsep}{1pt} \setlength{\parskip}{0pt} \setlength{\parsep}{0pt}}
	{\end{itemize} \vspace{-4pt}}
\newenvironment{enumerate*}[1][]
	{\vspace{-6pt} \begin{enumerate}[#1] \setlength{\itemsep}{1pt} \setlength{\parskip}{0pt} \setlength{\parsep}{0pt}}
	{\end{enumerate} \vspace{-4pt}}

\begin{document}

%\title{Identifying Key Cyber-Physical Terrain for Mobile Tactical Networks}
\title{Identifying Key Cyber-Physical Terrain (Extended Version)}
%\\ {\large (\today~~@~~\currenttime)}%}
%\\ {\LARGE\bfseries (Work in progress. Please do not distribute.)}}

%\numberofauthors{2}
%\author{
%\alignauthor Brian Thompson\\
	%\affaddr{The MITRE Corporation \& Army Research Lab}\\
	%\affaddr{7515 Colshire Dr.}\\
	%\affaddr{McLean, VA 22102}\\
	%\email{bthompson@mitre.org}
%\alignauthor Richard Harang\\
	%\affaddr{Invincea Inc. \& Army Research Lab}\\
	%\affaddr{3975 University Dr. \# 460}\\
	%\affaddr{Fairfax, VA 22030}\\
	%\email{rich.harang@gmail.com}
%}

\author{
	Brian Thompson\\
	The MITRE Corporation \& U.S. Army Research Lab\\
	7515 Colshire Dr.\\
	McLean, VA 22102\\
	bthompson@mitre.org
\and
	Richard Harang\\
	Invincea Inc. \& U.S. Army Research Lab\\
	3975 University Dr. \# 460\\
	Fairfax, VA 22030\\
	rich.harang@gmail.com
}

%keywords: risk, virus, worm, cyber attack, distributed network, agility

\maketitle

\begin{abstract}
%There is an inherent interconnectedness between the cyber and physical layers of tactical operations. Control over the cyber domain means the potential for better communication, more information about the enemy, less information leakage, which is likely to translate into greater control over the physical domain. This in turn means that it is easier to achieve cybersecurity and launch more effective cyber attacks. In this work, we explore that relationship(??) (Actually, only part of it is represented: the effect of physical control over cyber control. Direction for future work?)
The high mobility of Army tactical networks, combined with their close proximity to hostile actors, elevates the risks associated with short-range network attacks. The connectivity model for such short range connections under active operations is extremely fluid, and highly dependent upon the physical space within which the element is operating, as well as the patterns of movement within that space. To handle these dependencies, we introduce the notion of ``key cyber-physical terrain'': locations within an area of operations that allow for effective control over the spread of proximity-dependent malware in a mobile tactical network, even as the elements of that network are in constant motion with an unpredictable pattern of node-to-node connectivity. We provide an analysis of movement models and approximation strategies for finding such critical nodes, and demonstrate via simulation that we can identify such key cyber-physical terrain quickly and effectively.
\end{abstract}

%\keywords{mobile tactical networks; malware spread; cyber-physical systems}

\section{Introduction}
\label{sec:intro}

\subsection{Motivation}
\label{sec:intro-motivation}

Army tactical networks in the field face a unique set of security considerations not found in either more conventional wireless networks or fixed infrastructure networks. While much previous work in analyzing the spread of malware in networks (including tactical networks) focuses on the \emph{logical} connectivity of the graph over time, these logical connectivity paths are often dominated by long-range tactical links which introduce some degree of stability to the logical connectivity graph. However the close proximity of Army tactical networks to adversarial networks introduces new considerations in the form of \emph{spatial} properties of the network: which units are in close proximity to each other and at what times. This form of connectivity is particularly relevant in the case of attacks that restrict themselves to short-range wireless communications -- such as through 802.11 or Bluetooth network stacks -- which may be more difficult to detect due to their failure to cross more conventional security boundaries or higher-resource nodes capable of fielding more sophisticated intrusion detection systems.

The short range of these attacks means that -- at any given instant -- the communications graph available to the malware is effectively disconnected, and it is only the mobility of the infected components over time that brings new victims into range and allows it to propagate. In addition, detection or remediation of such infections may be prohibitively difficult to perform in the field, perhaps involving detailed scans, or simply complete reimaging or replacement of any potentially compromised devices, and so only carried out at particular locations. Furthermore, while standard defensive measures are effective against known malware and minor variants, novel (``zero-day'') attacks may be specifically developed for and deployed against military mobile networks. This malware may not be detectable, and so understanding how to bound the potential impact of such malware, even when not specifically alerted to its presence, is an important problem.

The notion of mobility over time, combined with the regularities in deployment and mobility of individual Army components (such as regular patrols, movement along roads and highways, and so forth), and limited capabilities to detect or remediate such attacks, leads us to our notion of \emph{key cyber-physical terrain}: critical points in the spatio-temporal graph which can be exploited to limit the spread of short-range malware. Identifying such critical points turns out to be surprisingly difficult in practice and so we explore several methods -- from simple graph-theoretic approaches to dynamical system approximations to full simulation -- that capture different aspects of this problem.

\subsection{Related Work}
\label{sec:intro-related}

Mathematical models of virus spread were first developed in the context of biological epidemics, primarily compartmental models which assume homogeneous interaction rates within the population, such as the well-known SIR (Susceptible-Infected-Recovered) model~\cite{kermack1927contribution} and its numerous variants. Kephart and White apply compartmental models to study the dynamics of malware spread in cyber networks, additionally using simulation to evaluate under which assumptions the compartmental models are most accurate~\cite{kephart1991directed}. They consider several network topologies, such as Erdos-Renyi random graphs, connected regular graphs, and sparse graphs with a high clustering coefficient. In all of these topological models they study the number of infected nodes in the population over time and how various factors affect convergence to a steady state, finding that in many cases there exists a sharp epidemic threshold. Others have extended such models to additional network topologies and contexts. For example, Boguna et al.~\cite{boguna2003epidemic} and Dezs\H{o} et al.~\cite{dezsHo2002halting} focus on epidemic models for power-law networks.

Marvel et al.\ propose a framework to evaluate cyber agility, but they focus on scenarios in which either a specific vulnerability or infected node is known to exist, and attempt to optimize the patching and isolation process in the network to preserve network integrity under various constraints including connectivity and power usage~\cite{marvel2015framework}. Huber et al.\ examine a similar problem using a decision support system in a small network of 10 active nodes~\cite{huber2016cyber}. Both cases assume malware with complete access to the network stack, which both allows longer-distance propagation than the local model we consider, and significantly increases the probability that the adversary will be detected.
%Furthermore, the defense mechanisms considered rely on effective detection capabilities. While detection is an important component of a comprehensive cyber defense strategy, it does little to protect against unknown vulnerabilities and zero-day attacks. We present an approach 

Mickens et al.\ study device-to-device spreading of malicious software in mobile ad-hoc networks (MANETs) by explicitly modeling node mobility~\cite{mickens2005modeling,mickens2007analytical}. Valler et al.\ develop a framework for analyzing malware spread in MANETs under the SIS (Susceptible-Infected-Susceptible) model~\cite{valler2011epidemic}. Su et al.\ perform simulations using trace data drawn from real-life sampling of over 10,000 devices in a commuter train station to examine the propagation dynamics of Bluetooth worms, showing that Bluetooth worms can infect a large population of vulnerable devices relatively quickly in an urban environment~\cite{su2006preliminary}. On the other hand, Wang et al.\ model the spread of malware across networks of mobile phone users and observe that Bluetooth-based malware spreads slowly due to the short range of Bluetooth and therefore the relatively low contact rate between devices~\cite{wang2009understanding}.
%Channakeshava et al.\ demonstrate via simulation on synthetic wireless networks using activity-based models of urban population mobility that the time it takes for a Bluetooth-based worm to spread throughout a network is highly dependent on the effectiveness of countermeasures taken within the first hour after it is introduced to the network~\cite{channakeshava2009epinet}.
%Other work has shown that malware spread in military contexts may differ significantly from that existing models
%make existing models unsuitable for modeling the propagation of malware in MTNs
This highlights the fact that the dynamics of malware spread in MANETs varies significantly based on the properties of the underlying movement patterns. In particular, the highly-structured movement often seen in military contexts differentiates mobile tactical networks from civil MANETs and impacts the propagation of malware in such settings~\cite{thompson2016impact}.
%In this work, we model tactical operations over a geographical region containing towns connected by a road network, framing the  as a resource allocation problem.
In this work, we explore how to leverage the structured mobility patterns of mobile tactical networks to develop more effective defense strategies, modeling tactical operations over a geographical region containing towns connected by a road network, and proposing computational methods to determine how to best allocate defensive resources.

\subsection{Contributions and Outline}
\label{sec:intro-contrib}

The main contributions of this work are:
\begin{itemize}
	\item Model and problem formulation highlighting the need for improved security in cyber-physical tactical operations
	%\item Model and problem formulation highlighting the need for improved security in cyber-physical tactical operations
	%\item A high-level model of tactical operations over a region consisting of towns connected by a road network
	\item Three computational approaches for deciding where to place remediation stations to best control the spread of malware
	\item Evaluation and comparison of the three approaches
	%\item Evaluation of the three strategies and comparison to optimal
\end{itemize}

In Section~\ref{sec:methods} we describe our tactical model and propose three computational approaches to determine the optimal defender strategy. In Section~\ref{sec:eval} we perform experiments to evaluate and compare the effectiveness of the approaches. We conclude with some discussion and directions for future work in Section~\ref{sec:conclusions}.

\section{Methods}
\label{sec:methods}

\subsection{Model and Problem Statement}
%\subsection{Tactical Model}
\label{sec:methods-model}

We consider a scenario in which tactical units of soldiers are deployed to towns in the same geographical region, connected by a road network. As time goes on, a unit may get redeployed to another town, at which point it
%traverses a shortest-path route
travels from its current town to the designated town through the road network.
%This could be implemented using the \emph{Graph Random Walk} or \emph{Graph Random Waypoint} movement pattern.

Each soldier is equipped with a mobile device that facilitates short-range wireless communication, such as Bluetooth, on the battlefield. Each device regularly scans the environment for nearby friendly devices. When two friendly devices come within communication range, they automatically connect, enabling data transmission.
%We assume that ...
%Due to its prevalence among mobile devices, we base our implementation on Bluetooth technology and will use that as our running example.

Enemy forces may attempt to infiltrate the allied cyber network by infecting allied devices with self-propagating malware, for example by infecting the device of a captured soldier or by deploying cyber hacking teams that can infect allied devices remotely. When a soldier with an infected device comes within range of a friendly soldier with an uninfected device, the malware spreads. The malware could, for example, give the enemy access to sensitive information, or the capability to corrupt data on infected devices.

%Some towns are controlled by the enemy and contain cyber hacking teams that attempt to infect the devices of allied units passing nearby. Other towns are under allied control and have been established as \emph{remediation zones}; any allied units entering such towns pass through a checkpoint where their devices are reset, replaced, or otherwise cleaned of malware.

To protect their cyber network from attack, allied forces may establish some towns as \emph{remediation zones}; any allied units entering such towns pass through a checkpoint where their devices are reset, replaced, or otherwise cleaned of malware. However, resources are limited, so judiciously choosing locations at which to establish remediation zones is critical.

\textbf{Objective:} Given knowledge of the road network, situational awareness of the location of enemy strongholds, and an assessment of remediation resources currently available, determine the optimal placement of remediation zones to minimize the fraction of devices that are infected with malware.

Below, we explore three approaches to addressing this problem: centrality analysis, dynamical systems, and agent-based modeling.

\subsection{Centrality Analysis}
\label{sec:methods-centrality}

In the centrality-based approach, we represent the road network as a graph and use network centrality analysis to identify the towns at which to establish remediation zones. The intuition is that the most central vertices are the most important, either visited most frequently or located at important junctures. Let $G$ be an undirected graph with vertex set $V(G) = \{v_1, \ldots, v_n\}$ corresponding to the towns and edge set $E(G) \subseteq {V \choose 2}$ corresponding to the roads. A centrality metric assigns weights to the vertices in a graph based on how central they are. For a given centrality metric $\mu$, we let $\mu_G : V(G) \to \R$ denote the mapping from the vertices of $G$ to their corresponding values under the centrality metric.

%intuition: the most frequently traveled towns

We consider two common centrality metrics:
%Degree centrality, PageRank centrality, and Betweenness centrality.
%natural correspondence between PageRank centrality and random walks
%natural correspondence between Betweenness centrality and random waypoint
\begin{itemize}
	%\item Degree centrality - favors vertices with many incident edges
	%, corresponding to towns with many roads connecting them to other towns
	%(note: if a road passes through a town, that counts twice)
	\item PageRank centrality~\cite{brin1998anatomy} - favors vertices with connections to other well-connected vertices
	%\item Eigenvector centrality - favors vertices based on properties of the adjacency matrix
	%\item Closeness centrality~\cite{?} - favors vertices that are close to many other vertices on average
	\item Betweenness centrality~\cite{freeman1977set} - favors vertices that lie on shortest paths between many other pairs of vertices
	%, corresponding to towns that will be passed through frequently
\end{itemize}
The choice of metric may be context-specific. For example, PageRank centrality has a natural correspondence with the frequency of vertices being visited under a mobility model where units perform a random walk on the road network, i.e. choosing the next town to visit uniformly at random from the set of neighboring towns. On the other hand, Betweenness centrality naturally corresponds with vertex frequency under a random waypoint mobility model, i.e. where units choose a town uniformly at random from the set of all towns and then traverse a shortest path to get there.

\begin{figure}[!t]
	\centering
		\includegraphics[height=1.5in]{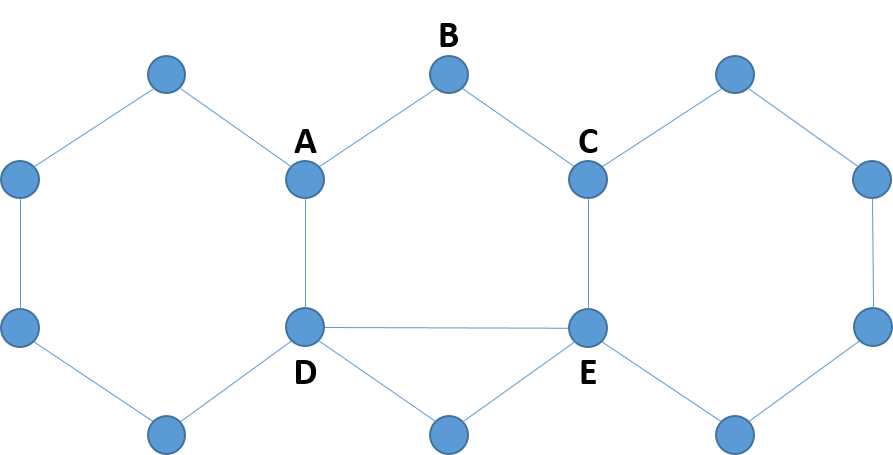}
	\caption{An example road network. Towns $D$ and $E$ may be the most central according to many metrics, but the best pair of towns would likely include one of $\{A,B,C\}$ and one of $\{D,E\}$.}
	\label{fig:centrality-counterexample-1}
\end{figure}

If there are only resources for a single remediation zone, centrality metrics offer a straight-forward way to choose where to place it: at the town corresponding to the vertex with the highest centrality score. If there are resources for $k > 1$ remediation zones, however, the natural solution of choosing the towns corresponding to the vertices with the $k$ highest centrality values may not be a very good strategy. For example, consider the graph in Figure~\ref{fig:centrality-counterexample-1} with $k = 2$. Vertices $D$ and $E$ have the top two centrality scores for PageRank and Betweenness centrality, yet a better strategy would likely be to choose one vertex in $\{A,B,C\}$ and one vertex in $\{D,E\}$ because that would cut the graph into two similarly-sized subgraphs between which malware could not propagate.

\begin{algorithm}[!t]
\caption{Centrality-based algorithm}
\label{alg:centrality}

\vspace{0.0625in}
\noindent {\bf Input}: A graph $G$, a centrality metric $\mu$, and an integer $k \geq 1$.

\vspace{0.0625in}
\noindent {\bf Output}: A subset $V_R^\ast \subseteq V(G)$ of size $k$ corresponding to the towns at which to establish remediation zones.

\begin{enumerate*}
	\item Initialize $G_0 := G$
	\item For $1 \leq i \leq k$:
		\begin{itemize}
			\item Compute $\mu_{i-1} = \mu_{G_{i-1}}$
			\item Set $v_i := \max\limits_{v \in V(G_{i-1})} \mu_{G_{i-1}}(v)$
			\item Set $G_i := G_{i-1} - v_i$
		\end{itemize}
	\item Return $V_R^\ast = \{v_i : 1 \leq i \leq k\}$
\end{enumerate*}

\end{algorithm}

To address this problem, we present an iterative algorithm, described in Algorithm~\ref{alg:centrality}. The algorithm computes centrality scores, deletes the vertex with the highest score, and repeats until $k$ vertices have been deleted. The remediation zones should be placed at the towns corresponding to the deleted vertices.

\begin{figure}[!t]
	\centering
		\includegraphics[height=1.1in]{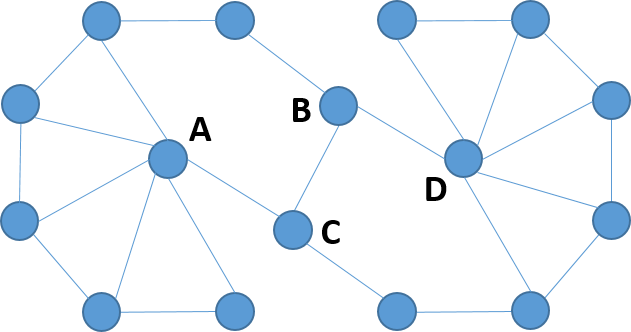}
	\caption{An example road network. The iterative centrality-based algorithm would remove $A$ and $D$ under many metrics, but the best pair of towns would likely include $B$ or $C$.}
	\label{fig:centrality-counterexample-2}
\end{figure}

However, there are still times when this does not produce the desired behavior. For example, consider the graph in Figure~\ref{fig:centrality-counterexample-2} with $k = 2$. Vertices $A$ and $D$ are tied for the top centrality score for PageRank and Betweenness centrality. After one of them is deleted, the other still has the highest score on the remaining graph. However, a better strategy would likely be to choose vertices $B$ and $C$ for the same reason as above.
%This is a drawback of the iterative algorithm, since the benefit of removing $B$ and $C$ is not apparent until both have been removed.

\begin{figure}[!t]
	\centering
		\includegraphics[height=1.5in]{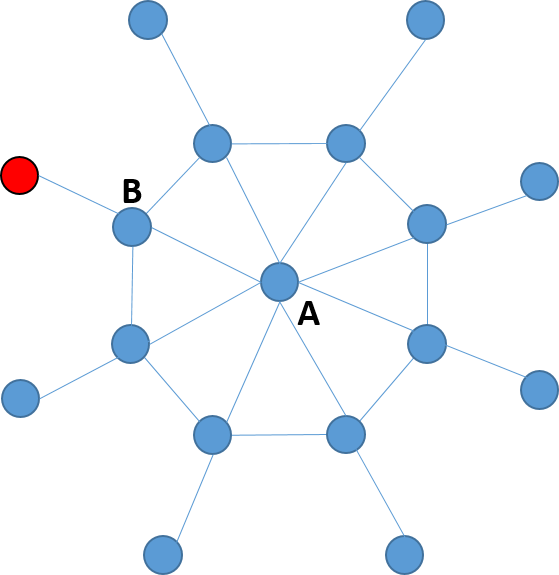}
	\caption{An example road network. Town $A$ may be the most central according to many metrics, but the best location for a remediation zone is at town $B$.}
	\label{fig:centrality-counterexample-3}
\end{figure}

Furthermore, there is no clear way of incorporating situational awareness of which towns are controlled by the enemy and therefore most likely that allied devices will get infected with malware. For example, consider the graph in Figure~\ref{fig:centrality-counterexample-3} with $k = 1$. Vertex $A$ has the highest centrality score for both metrics, but the best strategy would obviously be to choose vertex $B$. This is a drawback of any centrality-based algorithm, since they are based solely on the network topology and are not sensitive to the locations of enemy strongholds.

Next we consider an approach from the field of dynamical systems that addresses these problems.

\subsection{Dynamical Systems}
\label{sec:methods-markov}

In the dynamical systems approach, we begin by modeling the movement of each unit as a continuous-time Markov chain, where states correspond to towns and roads, and transitions correspond to changes in location in response to new tactical orders. When deployed at a town, a unit stays there for some \emph{deployment time} until it receives new orders to travel to a neighboring town. When it receives the travel order, it transitions to the road between the two towns, and remains there for the duration of the \emph{travel time}, which may depend on the distance, terrain, weather conditions, etc.

Let $S_i$ denote the state corresponding to town $i$, and let $S_{i,j}$ denote the state corresponding to traversing a road from town $i$ to town $j$. We define the average wait time $w_i$ for state $S_i$ to be equal to the average deployment time for town $i$. We define the average wait time $w_{i,j}$ for state $S_{i,j}$ to be equal to the average travel time from town $i$ to town $j$.

\begin{figure}[!t]
	\centering
	\includegraphics[width=0.3\textwidth]{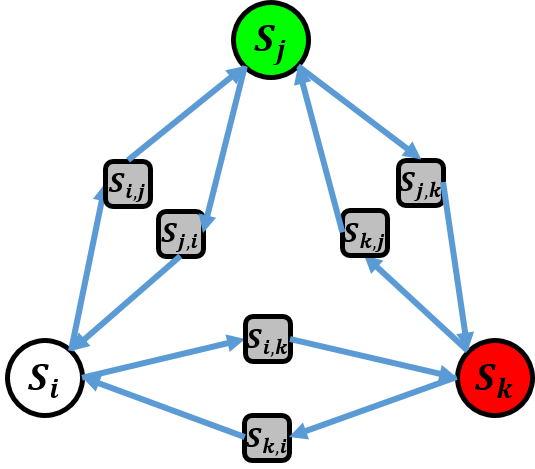}
	\caption{Markov chain for simple example scenario}
	\label{fig:model-dynsys-basic}
\end{figure}

%Let $T$ denote the transition matrix, i.e. $T(S,S')$ is the transition rate from state $S$ to state $S'$.
There are two types of transitions: from a state $S_i$ to a state $S_{i,j}$, corresponding to departure from town $i$ along a road to town $j$; and from a state $S_{i,j}$ to a state $S_j$, corresponding to arrival at town $j$ along a road from town $i$.
Assuming that units leaving a town have the same likelihood of traveling to each of the neighboring towns, the transition rates are as follows:
\begin{align*}
	(\forall\; i,j : (v_i,v_j) \in E(G)) \qquad T(S_i, S_{i,j}) \quad & = \quad \frac{1}{w_i \cdot d_i} \\
	(\forall\; i,j : (v_i,v_j) \in E(G)) \qquad T(S_{i,j}, S_j) \quad & = \quad \frac{1}{w_{i,j}}
\end{align*}
The Markov chain for a simple example scenario is illustrated in Figure~\ref{fig:model-dynsys-basic}.

%We now take a mean-field approach, modeling
Next, we describe the movement of all units collectively using a compartmental model corresponding to the Markov chain described above, %Let $\pi_i(t)$ denote the fraction of units that are in town $i$ at time $t$, and let $\pi_{i,j}(t)$ denote the fraction of units that are on the road from town $i$ to town $j$ at time $t$.
capturing the fraction of units in each state and the flows between them with a set of differential equations. These equations can then be used to solve for the fraction of units in each state at equilibrium, indicating which towns will be most frequently visited, which could be good candidates for remediation zones. A more detailed technical description of this approach is provided in Appendix A.

However, the same problems encountered under the centrality-based approach above still remain: choosing a set of towns based on each town's individual value may not yield the best results collectively; and we have not leveraged knowledge of the location of enemy strongholds.

\begin{figure}[!t]
	\centering
	\includegraphics[width=0.35\textwidth]{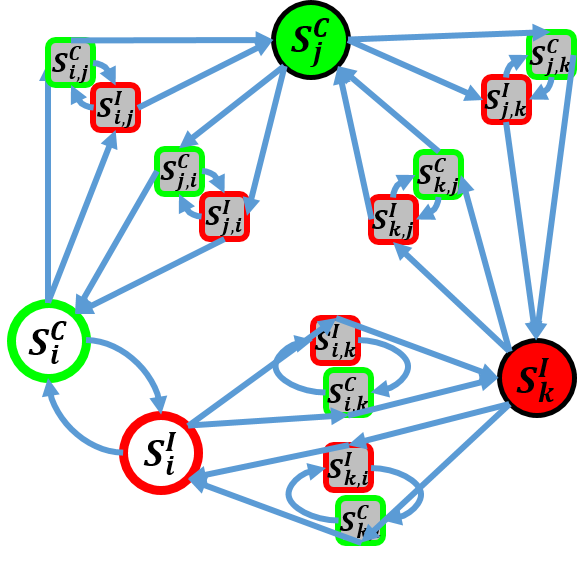}
	\caption{Modified Markov chain for simple example scenario}
	\label{fig:model-dynsys-advanced}
\end{figure}

To address these problems, we consider a modified Markov model that splits each previous state into two dual states, corresponding to whether the unit is clean or infected. We denote this by states $S_i^C$, $S_i^I$, $S_{i,j}^C$, and $S_{i,j}^I$.
%with limiting probabilities $\pi_i^C$, $\pi_i^I$, $\pi_{i,j}^C$, and $\pi_{i,j}^I$, respectively.
Let $V_I$ denote the set of vertices corresponding to enemy strongholds, and let $V_R$ denote the set of vertices corresponding to towns with remediation stations, with $V_R \cap V_I = \emptyset$. We assume that any unit entering a town in $V_I$ will become infected with malware, and any unit entering a town in $V_R$ will become clean. In addition, we assume that a clean unit entering a town with at least one infected unit will become infected, and also that a clean unit traversing a road with at least one infected unit traveling in the opposite direction will become infected. The goal is to determine the optimal set $V_R$ of size $k$, given $V_I$.
The modified Markov chain for the example scenario is illustrated in Figure~\ref{fig:model-dynsys-advanced}.

Similarly to above, the movement of all units collectively can be captured by a set of differential equations, which, given $V_I$ and $V_R$, can be solved efficiently for the equilibrium fraction of units in each state (see Appendix A for details). In this modified model, however, we have a way of quantifying the effectiveness of a proposed solution: the total fraction of infected units at equilibrium. The remaining challenge is in finding the set $V_R$ that minimizes that value.

\begin{algorithm}[!t]
\caption{Monte Carlo algorithm}
\label{alg:sampling}

\vspace{0.0625in}
\noindent {\bf Input}: A graph $G$ representing towns connected by a road network, a subset of vertices $V_I \subseteq V(G)$ corresponding to enemy strongholds, a function $f : 2^V(G) \to \R$ mapping vertex subsets $V_R$ to the resulting fraction of infected units if remediation zones were established at the corresponding towns, an integer $s$ indicating how many random samples to take at the corresponding point in the algorithm, and an integer $k \geq 1$ indicating the number of remediation zones for which resources are available.

\vspace{0.0625in}
\noindent {\bf Output}: A subset $V_R^\ast \subseteq V(G)$ of size $k$ corresponding to the towns at which to establish remediation zones.

%1. sample all
%2. take best solution found (but always consider the best found in the previous round)
%3. for each element of $V_R$, take the average over all solutions tried that include it
%4. choose the one with the highest average

\begin{enumerate}
	\item Initialize $V_R^{(0)} := \emptyset$  % current partial result
	\item Initialize $V_R^\ast := \emptyset$  % best found
	\item Initialize $f^\ast := 0$  % best found
	\item For $1 \leq i \leq k$:
		\begin{itemize}
			\item For $v \in V(G) - V_R^{(i-1)}$:
				\begin{itemize}
					\item Initialize $F^\text{sum}[v] = 0$
					\item Initialize $F^\text{count}[v] = 0$
				\end{itemize}
			\item for $1 \leq j \leq s$
				\begin{itemize}
					\item Randomly select a subset $V' \subseteq V(G) - V_R^{(i-1)}$ of size $k - (i-1)$
					\item Set $V'' := V_R^{(i-1)} \cup V'$
					\item For $v \in V'$:
						\begin{itemize}
							\item Update $F^\text{sum}[v] := F^\text{sum}[v] + f(V'')$
							\item Update $F^\text{count}[v] := F^\text{count}[v] + 1$
						\end{itemize}
					\item If $f(V'') < f^\ast$:
						\begin{itemize}
							\item Set $V_R^\ast := V''$
							\item Set $f^\ast := f(V'')$
						\end{itemize}
				\end{itemize}
			\item Set $v_i := \min\limits_{v \in V_R^\ast - V_R^{(i-1)}} \frac{F^\text{sum}[v]}{F^\text{count}[v]}$
			\item Set $V_R^{(i)} := V_R^{(i-1)} \cup \{v_i\}$
		\end{itemize}
	\item Return $V_R^\ast = V_R^{(k)} = \{v_i : 1 \leq i \leq k\}$
\end{enumerate}

\end{algorithm}

If $n$, the number of towns, and $k$, the desired number of remediation stations, are small, then an exhaustive search may be feasible. Otherwise, we propose two algorithms: one which simply entails sampling from the space of possible solutions and choosing whichever solution gives the best result; and one which is based on more sophisticated random sampling and Monte Carlo methods, described in Algorithm~\ref{alg:sampling}.

The dynamical systems approach addresses some of the major problems with the centrality-based approach, viz. considering
%the efficacy of establishing remediation stations at 
multiple towns simultaneously, and explicitly representing the presence of enemy strongholds. However, it is less flexible than the centrality-based approach in accommodating different mobility models; the Markov property is fine for modeling a random walk on the road network, but cannot easily represent multi-hop paths such as traversing the shortest path between two towns. In addition, the model makes several simplifying assumptions that could compromise the accuracy of the results.

Next, we present an approach that gives greater flexibility in modeling and also permits a higher degree of realism.

\subsection{Agent-based Modeling}
\label{sec:methods-abm}

\begin{figure}[!t]
	\centering
	\includegraphics[width=0.3\textwidth]{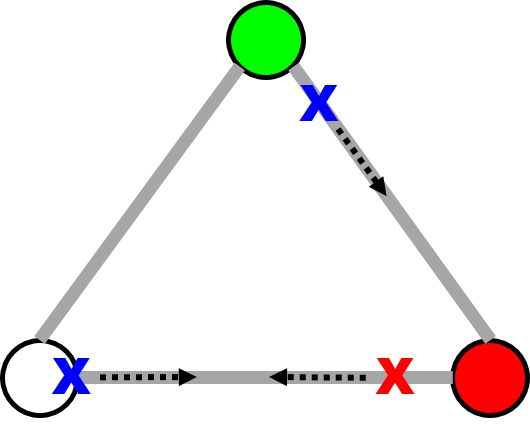}
	\caption{Agent-based model for simple example scenario}
	\label{fig:model-abm}
\end{figure}

In this approach, we develop an agent-based model to represent the movement of and interactions between tactical units.
%, based on the framework presented in [reference omitted for blind submission]. %~\cite{}.
%For simplicity, and to increase simulation speed, we model the units as atomic entities, instead of modeling at the granularity of individual soldiers.
The agents are the tactical units, each represented by a \tt{Unit} object. The environment consists of \tt{Town} objects, represented by circular regions, and \tt{Road} objects, each connecting two \tt{Town}s. \tt{Town}s can be ally-controlled, enemy-controlled, or neutral. This approach can accommodate many different mobility models, including both the random walk and the random waypoint models for traversing the road network. An agent-based model for a simple example scenario is illustrated in Figure~\ref{fig:model-abm}.

When a \tt{Unit} is deployed at or passes through an enemy-controlled \tt{Town}, we make the worst-case assumption that the enemy will be able to infect at least one of the soldiers' devices, and that relatively soon thereafter the malware will spread to the whole \tt{Unit} as the soldiers interact with one another. In addition, we assume that if two \tt{Unit}s are deployed to the same \tt{Town} simultaneously, or if one \tt{Unit} passes through the \tt{Town} where the other is deployed, or if two \tt{Unit}s pass each other on a \tt{Road}, there will be at least some contact between the \tt{Unit}s; therefore, if one of them is infected, the other will also become infected.

As before, our goal is to determine the set of vertices $V_R$ at which to place remediation zones so as to minimize the fraction of infected units. An obvious way to evaluate a proposed solution, then, is to run the simulation for a period of time and then count how many of the units are infected. Because of random variation, the result should be averaged over multiple trials. As with the dynamical systems approach, the remaining challenge is in finding the set $V_R$ that minimizes that value. For this, we propose using either the simple random sampling method or the same Monte Carlo algorithm proposed above, Algorithm~\ref{alg:sampling}, substituting the results of the agent-based simulation for the solution to the dynamical system when defining the function $f$.

The agent-based modeling approach has higher fidelity and expressiveness than the other approaches, but can also be more computationally expensive. In the following section, we evaluate both the effectiveness and computational efficiency of the three methods in determining the placement of remediation stations to best limit the spread of malware.

\section{Evaluation}
\label{sec:eval}

We now perform experiments to evaluate and compare the performance of the three approaches.
%in determining remediation zone placement strategies to control the spread of malware.
Since the agent-based model has the highest fidelity of the three approaches, we use it as an evaluative metric to compare different recommended placement strategies.
%That is, if we were to perform an exhaustive search of all possible placement strategies, we take as an assumption that the results of simulation using the agent-based model on each strategy would be the most accurate predictor of how they would perform in real life. However, as we will observe, that may not be computationally feasible.
Given that, one might expect that the agent-based modeling approach would trivially yield the best results. However, as we will observe, due to computational limitations this is not always the case.

Before we proceed with the experiments, we provide details of our implementation.
%provide details of our implementations of the three approaches and the simulation environment.
%describe the procedure we use for generating realistic-looking road networks.

\subsection{Implementation and Experimental Setup}

All three of the approaches are implemented in Java. Solving systems of equations for the dynamical systems approach was done using the JAMA linear algebra package. Simulations of the agent-based model can be visualized using Repast Simphony, a Java-based agent-based modeling and simulation environment. Experiments are conducted on an Intel Core i7 processor operating at 2.40 GHz with 16 GB of memory running Windows 10.

\begin{figure}[!t]
	\centering
	\includegraphics[width=0.6\textwidth]{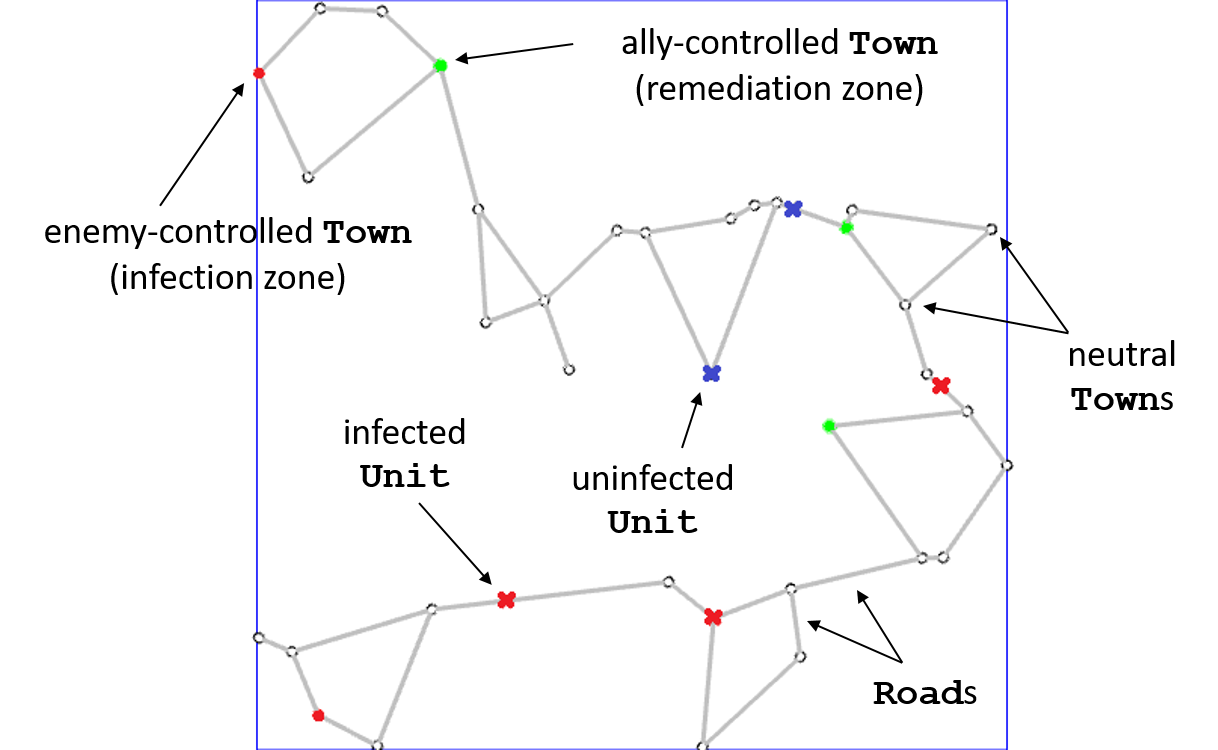}
	\caption{Labeled screenshot of agent-based simulation in Repast Simphony}
	\label{fig:sim-screenshot-roads}
\end{figure}

Figure~\ref{fig:sim-screenshot-roads} gives a screenshot of an example run of the agent-based simulation. The black circles represent neutral \tt{Town}s, the red circles represent \tt{Town}s under enemy control, and the green circles represent \tt{Town}s under allied control. \tt{Unit}s are depicted by a red `X' when infected and a black `X' when uninfected.

%\subsection{Experimental Setup}
%\label{sec:eval-setup}

For the experiments presented here, we consider five tactical units operating in a geographical area consisting of 35 towns connected by a road network. Units move at a speed of 10 m/s, and deployments last 2 hours. We vary the number of infected towns and remediation zones.
%consider three infected towns and three remediation zones.
%one infected town, and 1, 2, or 3 remediation zones.
%Mobile devices have a transmission range of 9.0 meters, a typical range for a Bluetooth signal in an unobstructed environment.
%
Simulations were run for 10,000 time steps.
%, corresponding to a period of 12 hours to represent the activities of a typical day, with data collected at time intervals of $\Delta t = 12$ seconds.
Results were averaged over 20 independent trials.

%high variance of (small samples of) agent-based sim

%[stochastic mobility model: Random Walk (and Random Waypoint?) on Graph]
%evaluative metrics: fraction of units infected, run-time
%how to measure success: fewest infected nodes, smallest spatial region containing infected nodes?, run-time

\subsection{Results}
\label{sec:eval-results}

\newcommand{\tblbrk}[2][c]{\begin{tabular}[#1]{@{}c@{}}#2\end{tabular}}

\begin{table*}[!t]
  \centering
  \caption{Experimental results under the Random Walk mobility model, in terms of the fraction of units infected, averaged over 20 trials}
    \begin{tabular}{|cc|cc|cc|cc|cc|c|}
    \toprule
    \multirow{2}[2]{*}{\textbf{\tblbrk{\# Inf \\ Zones}}} & \multirow{2}[2]{*}{\textbf{\tblbrk{\# Rmd \\ Zones}}} & \multicolumn{2}{c|}{\textbf{Betweenness}} & \multicolumn{2}{c|}{\textbf{PageRank}} & \multicolumn{2}{c|}{\textbf{Dynam Sys}} & \multicolumn{2}{c|}{\textbf{Agent-based}} & \multirow{2}[2]{*}{\textbf{\tblbrk{Uniform \\ Random}}} \\
          &       & \textbf{Top-k} & \textbf{Iter} & \textbf{Top-k} & \textbf{Iter} & \textbf{Basic} & \textbf{MC} & \textbf{Basic} & \textbf{MC} &  \\
    \midrule
    5     & 0     & \cellcolor[rgb]{ .973,  .412,  .42} 1.000 & \cellcolor[rgb]{ .973,  .412,  .42} 1.000 & \cellcolor[rgb]{ .973,  .412,  .42} 1.000 & \cellcolor[rgb]{ .973,  .412,  .42} 1.000 & \cellcolor[rgb]{ .973,  .412,  .42} 1.000 & \cellcolor[rgb]{ .973,  .412,  .42} 1.000 & \cellcolor[rgb]{ .973,  .412,  .42} 1.000 & \cellcolor[rgb]{ .973,  .412,  .42} 1.000 & \cellcolor[rgb]{ .973,  .412,  .42} 1.000 \\
    5     & 1     & \cellcolor[rgb]{ .984,  .573,  .451} 0.844 & \cellcolor[rgb]{ .984,  .584,  .455} 0.834 & \cellcolor[rgb]{ .98,  .557,  .447} 0.861 & \cellcolor[rgb]{ .984,  .573,  .451} 0.844 & \cellcolor[rgb]{ .984,  .624,  .463} 0.795 & \cellcolor[rgb]{ .984,  .624,  .463} 0.795 & \cellcolor[rgb]{ .98,  .545,  .447} 0.870 & \cellcolor[rgb]{ .98,  .545,  .447} 0.870 & \cellcolor[rgb]{ .98,  .557,  .447} 0.862 \\
    5     & 3     & \cellcolor[rgb]{ .992,  .718,  .478} 0.701 & \cellcolor[rgb]{ .992,  .71,  .478} 0.710 & \cellcolor[rgb]{ .992,  .761,  .49} 0.658 & \cellcolor[rgb]{ .996,  .839,  .502} 0.582 & \cellcolor[rgb]{ .996,  .851,  .506} 0.572 & \cellcolor[rgb]{ .996,  .847,  .506} 0.573 & \cellcolor[rgb]{ .992,  .745,  .486} 0.675 & \cellcolor[rgb]{ .996,  .78,  .494} 0.640 & \cellcolor[rgb]{ .988,  .702,  .478} 0.716 \\
    5     & 5     & \cellcolor[rgb]{ .996,  .808,  .498} 0.615 & \cellcolor[rgb]{ .996,  .804,  .498} 0.618 & \cellcolor[rgb]{ 1,  .91,  .518} 0.514 & \cellcolor[rgb]{ .922,  .898,  .51} 0.438 & \cellcolor[rgb]{ .925,  .898,  .51} 0.442 & \cellcolor[rgb]{ .906,  .894,  .51} 0.424 & \cellcolor[rgb]{ 1,  .882,  .51} 0.541 & \cellcolor[rgb]{ 1,  .91,  .518} 0.515 & \cellcolor[rgb]{ .996,  .827,  .502} 0.594 \\
    3     & 5     & \cellcolor[rgb]{ .973,  .914,  .514} 0.480 & \cellcolor[rgb]{ 1,  .894,  .514} 0.529 & \cellcolor[rgb]{ .886,  .886,  .51} 0.407 & \cellcolor[rgb]{ .749,  .847,  .502} 0.295 & \cellcolor[rgb]{ .796,  .863,  .506} 0.336 & \cellcolor[rgb]{ .722,  .839,  .498} 0.274 & \cellcolor[rgb]{ .929,  .902,  .514} 0.445 & \cellcolor[rgb]{ .867,  .882,  .51} 0.393 & \cellcolor[rgb]{ .988,  .918,  .514} 0.493 \\
    1     & 5     & \cellcolor[rgb]{ .631,  .812,  .494} 0.199 & \cellcolor[rgb]{ .69,  .831,  .498} 0.248 & \cellcolor[rgb]{ .616,  .808,  .494} 0.186 & \cellcolor[rgb]{ .525,  .784,  .49} 0.113 & \cellcolor[rgb]{ .541,  .788,  .49} 0.128 & \cellcolor[rgb]{ .51,  .78,  .486} 0.101 & \cellcolor[rgb]{ .627,  .812,  .494} 0.197 & \cellcolor[rgb]{ .525,  .784,  .49} 0.113 & \cellcolor[rgb]{ .714,  .839,  .498} 0.267 \\
    0     & 5     & \cellcolor[rgb]{ .388,  .745,  .482} 0.000 & \cellcolor[rgb]{ .388,  .745,  .482} 0.000 & \cellcolor[rgb]{ .388,  .745,  .482} 0.000 & \cellcolor[rgb]{ .388,  .745,  .482} 0.000 & \cellcolor[rgb]{ .388,  .745,  .482} 0.000 & \cellcolor[rgb]{ .388,  .745,  .482} 0.000 & \cellcolor[rgb]{ .388,  .745,  .482} 0.000 & \cellcolor[rgb]{ .388,  .745,  .482} 0.000 & \cellcolor[rgb]{ .388,  .745,  .482} 0.000 \\
    \bottomrule
    \end{tabular}%
  \label{tab:results-rwalk}%
\end{table*}%

% Table generated by Excel2LaTeX from sheet 'comparison'
\begin{table*}[!t]
  \centering
  \caption{Experimental results under the Random Waypoint mobility model, in terms of the fraction of units infected, averaged over 20 trials}
    \begin{tabular}{|cc|cc|cc|cc|cc|c|}
    \toprule
    \multirow{2}[2]{*}{\textbf{\tblbrk{\# Inf \\ Zones}}} & \multirow{2}[2]{*}{\textbf{\tblbrk{\# Rmd \\ Zones}}} & \multicolumn{2}{c|}{\textbf{Betweenness}} & \multicolumn{2}{c|}{\textbf{PageRank}} & \multicolumn{2}{c|}{\textbf{Dynam Sys}} & \multicolumn{2}{c|}{\textbf{Agent-based}} & \multirow{2}[2]{*}{\textbf{\tblbrk{Uniform \\ Random}}} \\
          &       & \textbf{Top-k} & \textbf{Iter} & \textbf{Top-k} & \textbf{Iter} & \textbf{Basic} & \textbf{MC} & \textbf{Basic} & \textbf{MC} &  \\
    \midrule
    5     & 0     & \cellcolor[rgb]{ .973,  .412,  .42} 1.000 & \cellcolor[rgb]{ .973,  .412,  .42} 1.000 & \cellcolor[rgb]{ .973,  .412,  .42} 1.000 & \cellcolor[rgb]{ .973,  .412,  .42} 1.000 & \cellcolor[rgb]{ .973,  .412,  .42} 1.000 & \cellcolor[rgb]{ .973,  .412,  .42} 1.000 & \cellcolor[rgb]{ .973,  .412,  .42} 1.000 & \cellcolor[rgb]{ .973,  .412,  .42} 1.000 & \cellcolor[rgb]{ .973,  .412,  .42} 1.000 \\
    5     & 1     & \cellcolor[rgb]{ .988,  .675,  .471} 0.743 & \cellcolor[rgb]{ .984,  .624,  .463} 0.793 & \cellcolor[rgb]{ .984,  .565,  .451} 0.852 & \cellcolor[rgb]{ .988,  .675,  .471} 0.743 & \cellcolor[rgb]{ .988,  .639,  .467} 0.779 & \cellcolor[rgb]{ .988,  .639,  .467} 0.779 & \cellcolor[rgb]{ .98,  .537,  .447} 0.878 & \cellcolor[rgb]{ .98,  .537,  .447} 0.878 & \cellcolor[rgb]{ .976,  .475,  .435} 0.940 \\
    5     & 3     & \cellcolor[rgb]{ 1,  .878,  .51} 0.543 & \cellcolor[rgb]{ .992,  .745,  .486} 0.673 & \cellcolor[rgb]{ .996,  .78,  .49} 0.640 & \cellcolor[rgb]{ .933,  .902,  .514} 0.446 & \cellcolor[rgb]{ 1,  .898,  .514} 0.526 & \cellcolor[rgb]{ 1,  .867,  .51} 0.556 & \cellcolor[rgb]{ .988,  .702,  .478} 0.717 & \cellcolor[rgb]{ .996,  .8,  .494} 0.621 & \cellcolor[rgb]{ .988,  .659,  .471} 0.758 \\
    5     & 5     & \cellcolor[rgb]{ .949,  .906,  .514} 0.461 & \cellcolor[rgb]{ .996,  .843,  .502} 0.581 & \cellcolor[rgb]{ 1,  .902,  .518} 0.520 & \cellcolor[rgb]{ .808,  .867,  .506} 0.345 & \cellcolor[rgb]{ .878,  .886,  .51} 0.403 & \cellcolor[rgb]{ .871,  .882,  .51} 0.395 & \cellcolor[rgb]{ 1,  .863,  .506} 0.561 & \cellcolor[rgb]{ .984,  .918,  .514} 0.489 & \cellcolor[rgb]{ .996,  .796,  .494} 0.624 \\
    3     & 5     & \cellcolor[rgb]{ .78,  .855,  .502} 0.321 & \cellcolor[rgb]{ .969,  .91,  .514} 0.474 & \cellcolor[rgb]{ .859,  .878,  .506} 0.385 & \cellcolor[rgb]{ .635,  .816,  .494} 0.203 & \cellcolor[rgb]{ .741,  .847,  .502} 0.289 & \cellcolor[rgb]{ .678,  .827,  .498} 0.238 & \cellcolor[rgb]{ .843,  .875,  .506} 0.373 & \cellcolor[rgb]{ .769,  .855,  .502} 0.312 & \cellcolor[rgb]{ .945,  .902,  .514} 0.455 \\
    1     & 5     & \cellcolor[rgb]{ .522,  .784,  .49} 0.112 & \cellcolor[rgb]{ .694,  .831,  .498} 0.252 & \cellcolor[rgb]{ .584,  .8,  .49} 0.160 & \cellcolor[rgb]{ .471,  .769,  .486} 0.068 & \cellcolor[rgb]{ .533,  .784,  .49} 0.122 & \cellcolor[rgb]{ .506,  .776,  .486} 0.096 & \cellcolor[rgb]{ .569,  .796,  .49} 0.148 & \cellcolor[rgb]{ .494,  .773,  .486} 0.088 & \cellcolor[rgb]{ .647,  .82,  .494} 0.214 \\
    0     & 5     & \cellcolor[rgb]{ .388,  .745,  .482} 0.000 & \cellcolor[rgb]{ .388,  .745,  .482} 0.000 & \cellcolor[rgb]{ .388,  .745,  .482} 0.000 & \cellcolor[rgb]{ .388,  .745,  .482} 0.000 & \cellcolor[rgb]{ .388,  .745,  .482} 0.000 & \cellcolor[rgb]{ .388,  .745,  .482} 0.000 & \cellcolor[rgb]{ .388,  .745,  .482} 0.000 & \cellcolor[rgb]{ .388,  .745,  .482} 0.000 & \cellcolor[rgb]{ .388,  .745,  .482} 0.000 \\
    \bottomrule
    \end{tabular}%
  \label{tab:results-rwaypoint}%
\end{table*}%

The results of our experiments are shown in Tables~\ref{tab:results-rwalk} (Random Walk mobility model) and~\ref{tab:results-rwaypoint} (Random Waypoint mobility model). For a baseline, we also record the average fraction of infected units when remediation zones are chosen uniformly at random.
%The considered methods performed similarly under the Random Walk and Random Waypoint mobility models;
%Due to space constraints, only the results for Random Waypoint are shown.

The best performers under the Random Walk mobility model were the Iterative PageRank Centrality method and the Dynamical Systems methods (either using simple random sampling or the more sophisticated Monte Carlo algorithm). The other methods performed significantly worse than those, and comparably to one another, sometimes not even matching the results of the uniform random baseline.

Under the more realistic Random Waypoint mobility model, Iterative PageRank was the clear winner, performing even better than under Random Walk. This was surprising, and ran counter to our intuition that PageRank would perform best under Random Walk because it has a natural correspondence to walks on graphs. Similarly, we were surprised that Betweenness centrality did not perform better under the Random Waypoint model, given its natural correspondence to graph paths. With the exception of the Iterative Betweenness method, all methods out-performed the baseline.

We note that we configured the ABM method to use fewer MC samples than the Dynamical Systems method (10 instead of 100) to keep its runtime comparable, since it does an evaluation over 10 sample trials for each candidate strategy rather than just solving a system of equations once.
%if we were to do it in another way that allowed 100 MC samples and/or 100 sample trials for ABM, I suspect it would do better. The 10 sample trials seems to really be a detriment, since with only 10 trials there is a lot of variance, so it might pick a solution that accidentally got some good trials instead of one that is consistently good.
We suspect that the small sample size resulted in a high variance across trials, which could explain why the ABM approach performed so poorly.

% Table generated by Excel2LaTeX from sheet 'comparison'
%\begin{table*}[!t]
  %\centering
  %\caption{Runtimes for the different approaches, in seconds}
    %\begin{tabular}{|c|cc|cc|cc|cc|cc|c|}
    %\toprule
    %\multirow{2}[2]{*}{} & \multicolumn{2}{c|}{\textbf{Betweenness}} & \multicolumn{2}{c|}{\textbf{PageRank}} & \multicolumn{2}{c|}{\textbf{Eigenvector}} & \multicolumn{2}{c|}{\textbf{Dynam Sys}} & \multicolumn{2}{c|}{\textbf{Agent-based}} & \textbf{Average} \\
          %& \textbf{Naive} & \textbf{Iter} & \textbf{Naive} & \textbf{Iter} & \textbf{Naive} & \textbf{Iter} & \textbf{Smpl} & \textbf{MC} & \textbf{Smpl} & \textbf{MC} & \textbf{Smpl} \\
    %\midrule
    %\textbf{Runtimes} & \cellcolor[rgb]{ 1,  .922,  .518} 0.194 & \cellcolor[rgb]{ .573,  .796,  .49} 0.190 & \cellcolor[rgb]{ .741,  .847,  .502} 0.192 & \cellcolor[rgb]{ .651,  .82,  .494} 0.191 & \cellcolor[rgb]{ .388,  .745,  .482} 0.188 & \cellcolor[rgb]{ .78,  .859,  .502} 0.192 & \cellcolor[rgb]{ .996,  .82,  .498} 0.311 & \cellcolor[rgb]{ .996,  .82,  .498} 0.308 & \cellcolor[rgb]{ .973,  .412,  .42} 0.761 & \cellcolor[rgb]{ .984,  .584,  .455} 0.570 & \cellcolor[rgb]{ .973,  .412,  .42} 0.761 \\
    %\bottomrule
    %\end{tabular}%
  %\label{tab:runtimes}%
%\end{table*}%

% Table generated by Excel2LaTeX from sheet 'comparison'
\begin{table*}[!t]
  \centering
  \caption{Runtimes for the different approaches, in seconds, averaged over 20 trials}
    \begin{tabular}{|c|cc|cc|cc|cc|c|}
    \toprule
    \multirow{2}[2]{*}{} & \multicolumn{2}{c|}{\textbf{Betweenness}} & \multicolumn{2}{c|}{\textbf{PageRank}} & \multicolumn{2}{c|}{\textbf{Dynam Sys}} & \multicolumn{2}{c|}{\textbf{Agent-based}} & \multirow{2}[2]{*}{\textbf{\tblbrk{Uniform \\ Random}}} \\
          & \textbf{Top-k} & \textbf{Iter} & \textbf{Top-k} & \textbf{Iter} & \textbf{Basic} & \textbf{MC} & \textbf{Basic} & \textbf{MC} &  \\
    \midrule
    \textbf{Runtimes} & \cellcolor[rgb]{ .388,  .745,  .482} 0.209 & \cellcolor[rgb]{ .388,  .745,  .482} 0.209 & \cellcolor[rgb]{ .388,  .745,  .482} 0.209 & \cellcolor[rgb]{ .388,  .745,  .482} 0.208 & \cellcolor[rgb]{ .627,  .812,  .494} 0.414 & \cellcolor[rgb]{ .635,  .816,  .494} 0.418 & \cellcolor[rgb]{ .973,  .412,  .42} 1.247 & \cellcolor[rgb]{ .984,  .616,  .459} 1.040 & \cellcolor[rgb]{ .973,  .412,  .42} 1.247 \\
    \bottomrule
    \end{tabular}%
  \label{tab:runtimes}%
\end{table*}%

Runtimes are shown in Table~\ref{tab:runtimes}. For this setting of the parameters, the centrality algorithms each ran in about 12 seconds, Dynamical Systems ran in about 25 seconds, ABM with MC in about 60 seconds, and ABM with random sampling in about 75 seconds.

\section{Conclusions}
\label{sec:conclusions}

\begin{table}[t]
	\centering
		\begin{tabular}{|L{0.75in}|L{1in}|L{1in}|}
			\hline
			Approach & Pros & Cons \\
			\hline
			Centrality metrics & can be efficient, choice of metric can accommodate different contexts or mobility patterns & does not capture travel times, cannot specify enemy towns, not good for multi-site selection \\
			\hline
			Dynamical systems & efficient, good for multi-site selection & assumes Random Walk mobility pattern because of Markov property \\
			\hline
			Agent-based modeling & very flexible and expressive, most realistic, good for multi-site selection & not as efficient as other approaches \\
			\hline
		\end{tabular}
	\caption{Pros and cons of the three approaches}
	\label{tab:pros-and-cons}
\end{table}
% efficiency, expressiveness, context-specific, multi-site selection

We have proposed the notion of ``key cyber-physical terrain'' to describe the risk posed by short-range wireless attacks under the dynamic connectivity graphs of field operations: specific physical locations at which mobile devices can be examined and remediated to minimize the ability of an adversary to maintain a presence on the mobile network. As the exact solution to this problem is computationally intractable, we have also proposed three approximate methods of solving the associated minimization problem -- centrality metrics, dynamical systems, and agent-based modeling -- under two different models of unit mobility. Some of their pros and cons are listed in Table~\ref{tab:pros-and-cons}.

Our results suggest that the problem of malware propagating through short-range wireless communications is potentially quite significant, with a high prevalence of malware persisting on the network, even when the remediation zones are placed strategically in response to the locations of the infection zones. It is also worth noting that simple algorithms based on network centrality metrics, in particular PageRank centrality, can match and even outperform more complex approximations, even under the more realistic Random Waypoint mobility model. In either case, we obtain solutions reasonably quickly, with average runtimes of about 1 minute even for our most computationally intensive approach. We note, however, that the variance for the agent-based modeling is relatively high, as the total number of potential trajectories through the combinatorial number of remediation zones is prohibitively large; results could be improved by increasing sample sizes in the Monte Carlo algorithm, at the cost of longer runtimes, which would be further exacerbated as the problem scales up. Methods to mitigate this variance will be explored in future work.
%As the problem scales up, we expect that the simpler approaches will provide more consistent results, and 

%We note that the exact numerical results are not quantitatively meaningful, since they are highly dependent on parameters such as the structure of the road network and the number of tactical units present.

%Assumption of ideal conditions, where the defender knows the attacker's locations (enemy strongholds).

Our current results show that both our centrality and dynamical systems methods can approach the accuracy of the more computationally intensive agent-based modeling approach under the mobility models used. On the other hand, the agent-based approach provides much greater flexibility for representing more sophisticated and realistic movement patterns and higher-fidelity models. For example, instead of random deployments and shortest-path traversals, simulations could be performed using real-world maps and scenarios, and paths may intentionally avoid locations of enemy strongholds. An alternative problem formulation could allow strategies to simultaneously define the traversal paths between pairs of towns as well as the locations of the remediation zones. This will be explored in future work, as well as extensions to our tactical model in which enemy infection regions as well as remediation zones may be dynamic or increase in number.

\bibliographystyle{abbrv}
\bibliography{../BIB/mobile-worm}

\appendix

\newpage
\newgeometry{margin=0.75in}

\section*{Appendix A: Theoretical Analysis of the Dynamical Systems Approach}
\label{sec:appendix-theory}

We consider $N$ tactical units moving along a road network connecting a set of $n$ towns. We represent the network as a directed graph $G$ with vertex set $V(G) = \{v_1, \ldots, v_n\}$ corresponding to the towns and edge set $E(G) \subseteq {V(G) \choose 2}$ corresponding to the roads.
%$E(G) \subseteq V(G) \times V(G)$ corresponding to the roads.
%The roads are directed to capture the fact that a unit typically travels in only one of the two possible directions at a given time.
Let $d_i$ denote the degree of vertex $v_i$ in $G$, corresponding to the number of roads into town $i$. Table~\ref{tab:notation} provides a summary of the notation used in our model.

\begin{table}[t]
	\centering
		\begin{tabular}{c|l}
			$N$ & the number of units/companies \\
			$n$ & the number of towns \\
			$v_i$ & vertex corresponding to town $i$ \\
			$(v_i,v_j)$ & edge corresponding to a road between town $i$ and town $j$ \\
			$d_i$ & the degree of $v_i$ (equivalently, the number of roads into town $i$) \\
			$S_i$ & state corresponding to being in town $i$ \\
			$S_{i,j}$ & state corresponding to being on the road from $i$ to $j$ \\
			$w_i$ & the average wait time for state $S_i$ (average deployment time at town $i$) \\
			$w_{i,j}$ & the average wait time for state $S_{i,j}$ (average travel time on the road from $i$ to $j$) \\
			$T$ & the transition matrix \\
			$\overline{\pi}_i$ & limiting probability of being in state $S_i$ \\
			$\overline{\pi}_{i,j}$ & limiting probability of being in state $S_{i,j}$ \\
			$S_i^C$, $S_i^I$ & states corresponding to being in town $i$ and clean/infected \\
			$S_{i,j}^C$, $S_{i,j}^I$ & states corresponding to being on the road from $i$ to $j$ and clean/infected \\
			$\overline{\pi}_i^C$, $\overline{\pi}_i^I$ & limiting probabilities of being in states $S_i^C$ and $S_i^I$, respectively \\
			$\overline{\pi}_{i,j}^C$, $\overline{\pi}_{i,j}^I$ & limiting probabilities of being in states $S_{i,j}^C$ and $S_{i,j}^I$, respectively
		\end{tabular}
	\caption{Notation used in our model.}
	\label{tab:notation}
\end{table}

We model the movement of each unit as a continuous-time Markov chain, where states correspond to towns and roads, and transitions correspond to changes in location in response to new tactical orders. When deployed at a town, a unit stays there for some \emph{deployment time} until it receives new orders to travel to a neighboring town. When it receives the travel order, it transitions to the road between the two towns, and remains there for the duration of the \emph{travel time}, which may depend on the distance, terrain, weather conditions, etc.

Let $S_i$ denote the state corresponding to town $i$, and let $S_{i,j}$ denote the state corresponding to traversing a road from town $i$ to town $j$. We define the average wait time $w_i$ for state $S_i$ to be equal to the average deployment time for town $i$. We define the average wait time $w_{i,j}$ for state $S_{i,j}$ to be equal to the average travel time from town $i$ to town $j$.

Let $T$ denote the transition matrix, i.e. $T(S,S')$ is the transition rate from state $S$ to state $S'$. There are two types of transitions: from a state $S_i$ to a state $S_{i,j}$, corresponding to departure from town $i$ along a road to town $j$; and from a state $S_{i,j}$ to a state $S_j$, corresponding to arrival at town $j$ along a road from town $i$. The transition rates are as follows:
\begin{align*}
	(\forall\; i,j : (v_i,v_j) \in E(G)) \qquad T(S_i, S_{i,j}) \quad & = \quad \frac{1}{w_i \cdot d_i} \\
	(\forall\; i,j : (v_i,v_j) \in E(G)) \qquad T(S_{i,j}, S_j) \quad & = \quad \frac{1}{w_{i,j}}
\end{align*}

We now take a mean-field approach, modeling all units collectively using a compartmental model corresponding to the Markov chain described above. Let $\pi_i(t)$ denote the fraction of units that are in town $i$ at time $t$, and let $\pi_{i,j}(t)$ denote the fraction of units that are on the road from town $i$ to town $j$ at time $t$. We provide the master equations expressing the instantaneous rate of change for each of the compartments:
\begin{align*}
	(\forall\; i : v_i \in V(G)) \qquad \frac{d \pi_i(t)}{dt} \quad & = \quad \(\sum_{j : (v_j,v_i) \in E(G)} \frac{\pi_{j,i}(t)}{w_{j,i}}\) - \frac{\pi_i(t)}{w_i} \\
	(\forall\; i,j : (v_i,v_j) \in E(G)) \qquad \frac{d \pi_{i,j}(t)}{dt} \quad & = \quad \frac{\pi_i(t)}{w_i \cdot d_i} - \frac{\pi_{i,j}(t)}{w_{i,j}}
\end{align*}

The limiting distribution over node states can be determined by solving the following system of equations, derived by setting $\frac{d \pi(t)}{dt} = 0$ for all states, where $\overline{\pi}_i$ and $\overline{\pi}_{i,j}$ are the limiting probabilities:
\begin{align*}
	(\forall\; i : v_i \in V(G)) \qquad & \frac{\overline{\pi}_i}{w_i} \quad = \quad \sum_{j : (v_j,v_i) \in E(G)} \frac{\overline{\pi}_{j,i}}{w_{j,i}} \\
	(\forall\; i,j : (v_i,v_j) \in E(G)) \qquad & \frac{\overline{\pi}_{i,j}}{w_{i,j}} \quad = \quad \frac{\overline{\pi}_i}{w_i \cdot d_i} \\
	& \sum_i \overline{\pi}_i + \sum_{i,j} \overline{\pi}_{i,j} \quad = \quad 1
\end{align*}
Since this is a system of $n + m + 1$ linear equations --- $n + m$ of which are linearly independent --- in $n + m$ variables, it can be solved efficiently, e.g. using Gaussian elimination.

Next we consider enemy cyber hacking teams planted in some of the towns. When an allied unit gets deployed to one of those towns, the enemy cyber team hacks into a soldier's device and infects it with a self-propagating Bluetooth worm, which then spreads to the devices of other soldiers in the unit as they continue carrying out their tactical objectives. In addition, malware then spreads from an infected unit to a clean unit when the two units are deployed in the same town at the same time, or when they pass each other on the road (going in opposite directions). We represent this with a modified Markov model that splits each previous state into two dual states, corresponding to whether the unit is clean or infected. We denote this by states $S_i^C$, $S_i^I$, $S_{i,j}^C$, and $S_{i,j}^I$.
%with limiting probabilities $\pi_i^C$, $\pi_i^I$, $\pi_{i,j}^C$, and $\pi_{i,j}^I$, respectively.
Let $V_I$ denote the set of vertices corresponding to towns with enemy cyber hacking teams.

For defensive strategy, we consider ``cleaning stations'' that may be placed at the entrance to towns. Any infected units that pass through one of the cleaning stations will become clean. Let $V_C$ denote the set of vertices corresponding to towns with cleaning stations. We assume that $V_C \cap V_I = \emptyset$.

There are now nine types of transitions:
\begin{itemize}
	\item from $S_i^C$ to $S_{i,j}^C$, when a clean unit departs from town $i$ along a road to town $j$, and there are currently no infected units traveling from town $j$ to town $i$
	\item from $S_{i,j}^C$ to $S_j^C$, when a clean unit arrives at town $j$ along a road from town $i$, and there are currently no infected units deployed at town $j$ and no enemy cyber hacking team
	\item from $S_i^C$ to $S_{i,j}^I$, when a clean unit departs from town $i$ along a road to town $j$, and there is currently an infected unit traveling from town $j$ to town $i$
	\item from $S_{i,j}^C$ to $S_j^I$, when a clean unit arrives at town $j$ along a road from town $i$, and there is currently an infected unit deployed at town $j$ or an enemy cyber hacking team
	\item from $S_i^C$ to $S_i^I$, when a clean unit gets infected in a town due to the arrival of an infected unit
	\item from $S_{i,j}^C$ to $S_{i,j}^I$, when a clean unit gets infected on a road by an infected unit starting to travel down the road in the opposite direction
	\item from $S_i^I$ to $S_{i,j}^I$, when an infected unit departs from town $i$ along a road to town $j$
	\item from $S_{i,j}^I$ to $S_j^I$, when an infected unit arrives at town $j$ along a road from town $i$, and there is not a cleaning station at town $j$
	\item from $S_{i,j}^I$ to $S_j^C$, when an infected unit arrives at town $j$ along a road from town $i$, and there is a cleaning station at town $j$
\end{itemize}

The corresponding transition rates are as follows, $\forall\; i : v_i \in V(G)$ and $\forall\; i,j : v_i,v_j \in V(G),\ (v_i,v_j) \in E(G)$:
\begin{align*}
	T(S_i^C, S_{i,j}^C)(t) \quad & = \quad \frac{\(1 - \pi_{j,i}^I(t)\)^{N-1}}{w_i \cdot d_i} \\
	T(S_{i,j}^C, S_j^C) \quad & = \quad
		\begin{dcases*}
			\frac{1}{w_{i,j}} & if $v_j \in V_C$ \\
			0 & if $v_j \in V_I$ \\
			\frac{\(1 - \pi_j^I(t)\)^{N-1}}{w_{i,j}} & otherwise
		\end{dcases*} \\
	T(S_i^C, S_{i,j}^I) \quad & = \quad \frac{1 - \(1 - \pi_{j,i}^I(t)\)^{N-1}}{w_i \cdot d_i} \\
	T(S_{i,j}^C, S_j^I) \quad & = \quad
		\begin{dcases*}
			0 & if $v_j \in V_C$ \\
			\frac{1}{w_{i,j}} & if $v_j \in V_I$ \\
			\frac{1 - \(1 - \pi_j^I(t)\)^{N-1}}{w_{i,j}} & otherwise
		\end{dcases*} \\
	T(S_i^C, S_i^I) \quad & = \quad (N - 1) \cdot \sum_{j} \frac{\pi_{j,i}^I(t)}{w_{j,i}} \\
	T(S_{i,j}^C, S_{i,j}^I) \quad & = \quad (N - 1) \cdot \frac{\pi_j^I(t)}{w_j \cdot d_j} \\
	T(S_i^I, S_{i,j}^I) \quad & = \quad \frac{1}{w_i \cdot d_i} \\
	T(S_{i,j}^I, S_j^I) \quad & = \quad
		\begin{dcases*}
			0 & if $v_j \in V_C$ \\
			\frac{1}{w_{i,j}} & otherwise
		\end{dcases*} \\
	T(S_{i,j}^I, S_j^C) \quad & = \quad
		\begin{dcases*}
			\frac{1}{w_{i,j}} & if $v_j \in V_C$ \\
			0 & otherwise
		\end{dcases*}
\end{align*}

This corresponds to a new set of master equations, also $\forall\; i : v_i \in V(G)$ and $\forall\; i,j : v_i,v_j \in V(G),\ (v_i,v_j) \in E(G)$:
\begin{align*}
	\frac{d \pi_i^C(t)}{dt} \quad & = \quad
		\begin{dcases*}
			\(\sum_{j : (v_j,v_i) \in E(G)} \frac{\pi_{j,i}^C(t) + \pi_{j,i}^I(t)}{w_{j,i}}\) - \frac{\pi_i^C(t)}{w_i} & if $v_i \in V_C$ \\
			0 & if $v_i \in V_I$ \\
			\(\sum_{j : (v_j,v_i) \in E(G)} \frac{\pi_{j,i}^C(t)}{w_{j,i}}\) \cdot \(1 - \pi_i^I(t)\)^{N-1} - \frac{\pi_i^C(t)}{w_i} & otherwise
		\end{dcases*} \\
	\frac{d \pi_i^I(t)}{dt} \quad & = \quad
		\begin{dcases*}
			0 & if $v_i \in V_C$ \\
			\(\sum_{j : (v_j,v_i) \in E(G)} \frac{\pi_{j,i}^C(t) + \pi_{j,i}^I(t)}{w_{j,i}}\) - \frac{\pi_i^I(t)}{w_i} & if $v_i \in V_I$ \\
			\begin{multlined}
				\(\sum_{j : (v_j,v_i) \in E(G)} \frac{\pi_{j,i}^C(t)}{w_{j,i}}\) \cdot \(1 - \(1 - \pi_i^I(t)\)^{N-1}\) + \(\sum_{j : (v_j,v_i) \in E(G)} \frac{\pi_{j,i}^I(t)}{w_{j,i}}\) \\
				+ (N-1) \cdot \pi_i^C(t) \cdot \sum_{j} \frac{\pi_{j,i}^I(t)}{w_{j,i}} - \frac{\pi_i^I(t)}{w_i}
			\end{multlined}
				& otherwise
		\end{dcases*} \\
	\frac{d \pi_{i,j}^C(t)}{dt} \quad & = \quad \frac{\pi_i^C(t)}{w_i \cdot d_i} \cdot \(1 - \pi_{j,i}^I(t)\)^{N-1} - \frac{\pi_{i,j}^C(t)}{w_{i,j}} \\
	\frac{d \pi_{i,j}^I(t)}{dt} \quad & = \quad \frac{\pi_i^C(t)}{w_i \cdot d_i} \cdot \(1 - \(1 - \pi_{j,i}^I(t)\)^{N-1}\) + \frac{\pi_i^I(t)}{w_i \cdot d_i} + (N-1) \cdot \pi_{i,j}^C(t) \cdot \frac{\pi_j^I(t)}{w_j \cdot d_j} - \frac{\pi_{i,j}^I(t)}{w_{i,j}}
\end{align*}

Setting $\frac{d \pi(t)}{dt} = 0$ for all states yields the following system of equations, where $\overline{\pi}_i^C$, $\overline{\pi}_i^I$, $\overline{\pi}_{i,j}^C$, and $\overline{\pi}_{i,j}^I$ are the limiting probabilities, again $\forall\; i : v_i \in V(G)$ and $\forall\; i,j : v_i,v_j \in V(G),\ (v_i,v_j) \in E(G)$:
\begin{align*}
	(\forall\; i : v_i \in V_C) \qquad & \left\{\ 
		\begin{aligned}
			& \frac{\overline{\pi}_i^C}{w_i} \quad = \quad \(\sum_{j : (v_j,v_i) \in E(G)} \frac{\overline{\pi}_{j,i}^C + \overline{\pi}_{j,i}^I}{w_{j,i}}\) \\
			& \frac{\overline{\pi}_i^I}{w_i} \quad = \quad 0
		\end{aligned} \right. \\
	(\forall\; i : v_i \in V_I) \qquad & \left\{\ 
		\begin{aligned}
			& \frac{\overline{\pi}_i^C}{w_i} \quad = \quad 0 \\
			& \frac{\overline{\pi}_i^I}{w_i} \quad = \quad \(\sum_{j : (v_j,v_i) \in E(G)} \frac{\overline{\pi}_{j,i}^C + \overline{\pi}_{j,i}^I}{w_{j,i}}\)
		\end{aligned} \right. \\
	(\forall\; i : v_i \notin V_C \cup V_I) \qquad & \left\{\ 
		\begin{aligned}
			& \frac{\overline{\pi}_i^C}{w_i} \quad = \quad \(\sum_{j : (v_j,v_i) \in E(G)} \frac{\overline{\pi}_{j,i}^C}{w_{j,i}}\) \cdot \(1 - \overline{\pi}_i^I\)^{N-1} \\
			& \frac{\overline{\pi}_i^I}{w_i} \quad = \quad
				\begin{multlined}
					\(\sum_{j : (v_j,v_i) \in E(G)} \frac{\overline{\pi}_{j,i}^C}{w_{j,i}}\) \cdot \(1 - \(1 - \overline{\pi}_i^I\)^{N-1}\) \\
					+ \(\sum_{j : (v_j,v_i) \in E(G)} \frac{\overline{\pi}_{j,i}^I}{w_{j,i}}\) + (N-1) \cdot \overline{\pi}_i^C \cdot \sum_{j} \frac{\overline{\pi}_{j,i}^I}{w_{j,i}}
				\end{multlined}
		\end{aligned} \right. \\
	(\forall\; i,j : (v_i,v_j) \in E(G)) \qquad & \left\{\ 
		\begin{aligned}
			& \frac{\overline{\pi}_{i,j}^C}{w_{i,j}} \quad = \quad \frac{\overline{\pi}_i^C}{w_i \cdot d_i} \cdot \(1 - \overline{\pi}_{j,i}^I\)^{N-1} \\
			& \frac{\overline{\pi}_{i,j}^I}{w_{i,j}} \quad = \quad \frac{\overline{\pi}_i^C}{w_i \cdot d_i} \cdot \(1 - \(1 - \overline{\pi}_{j,i}^I\)^{N-1}\) + \frac{\overline{\pi}_i^I}{w_i \cdot d_i} + (N-1) \cdot \overline{\pi}_{i,j}^C \cdot \frac{\overline{\pi}_j^I}{w_j \cdot d_j}
		\end{aligned} \right. \\
	& \sum_i \overline{\pi}_i^C + \sum_i \overline{\pi}_i^I + \sum_{i,j} \overline{\pi}_{i,j}^C + \sum_{i,j} \overline{\pi}_{i,j}^I \quad = \quad 1
\end{align*}

We note that some of these equations are non-linear. However, the non-linear terms can be approximated by substituting $\overline{\pi}_i$ or $\overline{\pi}_{i,j}$ --- whose values can be computed efficiently from the first system of equations ---
%~\ref{?}
for each occurrence of $\overline{\pi}^I_i$ or $\overline{\pi}^I_{i,j}$, respectively. This substitution over-estimates fractions of infected nodes, thus yielding a set of linear equations in the new variables $\undertilde{\overline{\pi}}$, with the property that $(\forall\; i)\ \overline{\pi}_i^I \leq \undertilde{\overline{\pi}}_i^I$ and $(\forall\; i,j)\ \overline{\pi}_{i,j}^I \leq \undertilde{\overline{\pi}}_{i,j}^I$:
%[proof using the concept of coupling (e.g. for percolation)?]
\begin{align*}
	(\forall\; i : v_i \in V_C) \qquad & \left\{\ 
		\begin{aligned}
			& \frac{\undertilde{\overline{\pi}}_i^C}{w_i} \quad = \quad \(\sum_{j : (v_j,v_i) \in E(G)} \frac{\undertilde{\overline{\pi}}_{j,i}^C + \undertilde{\overline{\pi}}_{j,i}^I}{w_{j,i}}\) \\
			& \frac{\undertilde{\overline{\pi}}_i^I}{w_i} \quad = \quad 0
		\end{aligned} \right. \\
	(\forall\; i : v_i \in V_I) \qquad & \left\{\ 
		\begin{aligned}
			& \frac{\undertilde{\overline{\pi}}_i^C}{w_i} \quad = \quad 0 \\
			& \frac{\undertilde{\overline{\pi}}_i^I}{w_i} \quad = \quad \(\sum_{j : (v_j,v_i) \in E(G)} \frac{\undertilde{\overline{\pi}}_{j,i}^C + \undertilde{\overline{\pi}}_{j,i}^I}{w_{j,i}}\)
		\end{aligned} \right. \\
	(\forall\; i : v_i \notin V_C \cup V_I) \qquad & \left\{\ 
		\begin{aligned}
			& \frac{\undertilde{\overline{\pi}}_i^C}{w_i} \quad = \quad \(\sum_{j : (v_j,v_i) \in E(G)} \frac{\undertilde{\overline{\pi}}_{j,i}^C}{w_{j,i}}\) \cdot \(1 - \overline{\pi}_i\)^{N-1} \\
			& \frac{\undertilde{\overline{\pi}}_i^I}{w_i} \quad = \quad
				\begin{multlined}
					\(\sum_{j : (v_j,v_i) \in E(G)} \frac{\undertilde{\overline{\pi}}_{j,i}^C}{w_{j,i}}\) \cdot \(1 - \(1 - \overline{\pi}_i\)^{N-1}\) \\
					+ \(\sum_{j : (v_j,v_i) \in E(G)} \frac{\undertilde{\overline{\pi}}_{j,i}^I}{w_{j,i}}\) + (N-1) \cdot \undertilde{\overline{\pi}}_i^C \cdot \sum_{j} \frac{\overline{\pi}_{j,i}}{w_{j,i}}
				\end{multlined}
		\end{aligned} \right. \\
	(\forall\; i,j : (v_i,v_j) \in E(G)) \qquad & \left\{\ 
		\begin{aligned}
			& \frac{\undertilde{\overline{\pi}}_{i,j}^C}{w_{i,j}} \quad = \quad \frac{\undertilde{\overline{\pi}}_i^C}{w_i \cdot d_i} \cdot \(1 - \overline{\pi}_{j,i}\)^{N-1} \\
			& \frac{\undertilde{\overline{\pi}}_{i,j}^I}{w_{i,j}} \quad = \quad \frac{\undertilde{\overline{\pi}}_i^C}{w_i \cdot d_i} \cdot \(1 - \(1 - \overline{\pi}_{j,i}\)^{N-1}\) + \frac{\undertilde{\overline{\pi}}_i^I}{w_i \cdot d_i} + (N-1) \cdot \undertilde{\overline{\pi}}_{i,j}^C \cdot \frac{\overline{\pi}_j}{w_j \cdot d_j}
		\end{aligned} \right. \\
	& \sum_i \undertilde{\overline{\pi}}_i^C + \sum_i \undertilde{\overline{\pi}}_i^I + \sum_{i,j} \undertilde{\overline{\pi}}_{i,j}^C + \sum_{i,j} \undertilde{\overline{\pi}}_{i,j}^I \quad = \quad 1
\end{align*}

\end{document}